\documentclass[10pt,letterpaper]{article}
\usepackage[letterpaper,margin=1in]{geometry}
\usepackage[T1]{fontenc}
\usepackage[utf8]{inputenc}
\usepackage{lmodern}
\usepackage{microtype}
\usepackage{authblk}
\usepackage{amsmath,amssymb}
\usepackage{booktabs}
\usepackage{array}
\usepackage{graphicx}
\usepackage{caption}
\usepackage{float}
\usepackage{xcolor}
\usepackage{placeins}
\usepackage{titlesec}
\titlespacing*{\section}{0pt}{12pt plus 2pt minus 2pt}{6pt}
\titlespacing*{\subsection}{0pt}{10pt plus 2pt minus 2pt}{4pt}
\titlespacing*{\paragraph}{0pt}{7pt plus 1pt minus 1pt}{1em}
\AtBeginDocument{%
  \setlength{\abovedisplayskip}{7pt plus 2pt minus 2pt}%
  \setlength{\belowdisplayskip}{7pt plus 2pt minus 2pt}%
  \setlength{\abovedisplayshortskip}{4pt plus 2pt}%
  \setlength{\belowdisplayshortskip}{4pt plus 2pt minus 2pt}%
}
\usepackage[numbers,sort&compress]{natbib}
\usepackage{hyperref}
\usepackage{doi}
\hypersetup{colorlinks=true,linkcolor=black,citecolor=black,urlcolor=black}
\title{stMIND: Learning multiscale interaction dynamics from spatial-temporal transcriptomics}

\author[1,2]{Longquan Li}

\author[1,2,*]{Lin Wan}

\affil[1]{School of Advanced Interdisciplinary Sciences, University of Chinese Academy of Sciences, Beijing, China.}
\affil[2]{Academy of Mathematics and Systems Science, Chinese Academy of Sciences, Beijing, China.}
\affil[*]{Corresponding author: \href{mailto:lwan@amss.ac.cn}{lwan@amss.ac.cn}}
\date{}

\begin{document}

\maketitle

\section*{Abstract}

Interactions within local cellular neighborhoods and across broader tissue scales jointly govern molecular state transitions and morphogenesis during development and regeneration. While time-resolved spatial transcriptomics simultaneously profiles gene expression and spatial coordinates, destructive sequencing samples distinct cells at each time point, leaving intermediate trajectories and multiscale intercellular interactions unobserved. Building upon our interacting mean-field neural stochastic differential equation framework, we present \textbf{stMIND} (\textbf{S}patiotemporal \textbf{M}ultiscale \textbf{I}nteraction \textbf{N}eural \textbf{D}ynamics), a continuous-time generative framework that infers coupled molecular, spatial, and proliferation dynamics from unmatched spatiotemporal snapshots. stMIND decomposes cellular expression velocity vector/drift into three interpretable operators: a cell-intrinsic program, a local graph-difference attention operator that aggregates relative expression differences and spatial geometry within a co-evolving physical radius, and a nonlocal mean-field attention operator that captures long-range population context beyond the immediate neighborhood. By evolving dynamic spatial neighborhoods alongside cellular migration and enforcing orthogonality between intrinsic and interaction-driven drifts, stMIND resolves scale-specific intercellular influences through optimal transport in joint expression--position space. Across benchmarks in axolotl brain regeneration and zebrafish embryogenesis, stMIND achieves the highest gene-expression prediction accuracy, anatomical consistency and spatial coherence. Ablation studies confirm that combining disjoint local and nonlocal operators is essential for accurately reconstructing cellular microenvironments. Furthermore, stMIND uncovers a critical regenerative transition at 10--15 days post-injury in the axolotl telencephalon—marked by a surge in total drift energy followed by nonlocal dominance and cell-type-specific local counterbalance within reactive ependymoglial niches. In addition, it accurately tracks stage-dependent developmental vector fields in zebrafish.

\par\noindent\hspace*{1em}\textbf{Code availability:} The code of stMIND is available at \url{https://github.com/LongquanLi135/stMIND}.

\clearpage

\section{Introduction}

\begingroup\color{black}
Tissue development and regeneration are orchestrated by dynamical processes operating across multiple spatial and temporal scales. Cell-intrinsic regulatory programs, short-range juxtacrine or paracrine signaling within immediate microenvironments, and longer-range morphogenetic cues across the broader tissue architecture jointly shape cellular states and spatial organization. As cells proliferate and migrate, the local and global cellular contexts experienced by each cell continuously co-evolve. Recent advances in spatially resolved transcriptomics enable simultaneous measurement of high-dimensional gene expression and physical tissue coordinates across developmental and regenerative stages  \cite{stahl2016,rao2021,vickovic2019}. However, because sequencing assays are destructive, time-series spatial transcriptomics captures static, unaligned population snapshots rather than longitudinal single-cell trajectories. A central computational challenge is therefore to reconstruct continuous coupled trajectories of gene expression and spatial migration from unmatched snapshots while disentangling how intercellular interactions across distinct spatial scales drive cell-state transitions \cite{armingol2020,armingol2024}.

Dynamic optimal transport (OT) and flow-based generative models have emerged as powerful paradigms for inferring continuous cellular trajectories from time-series single-cell data \cite{schiebinger2019,tong2020,huguet2022,yeo2021,jiang2024,zhang2024}. Extending these principles to time-series spatial transcriptomics, STORIES incorporates spatial geometry into an optimal-transport objective to infer cell-fate landscapes \cite{huizing2026stories}, while stVCR jointly models gene expression differentiation, proliferation, migration, and spatial coordinate alignment within a continuous optimal-transport framework \cite{peng2026}. Although these methods reconstruct aspects of spatiotemporal tissue evolution, their velocity vector/drift parameterizations do not explicitly model or disentangle local and nonlocal intercellular interactions in the governing expression dynamics.

Conversely, interaction-aware dynamical models explicitly incorporate cell--cell communication into continuous-time state equations. Recently, we introduced scIMF, which embeds cell-wise self-attention within an interacting mean-field (McKean--Vlasov) neural stochastic differential equation (SDE) to learn collective multicellular dynamics \cite{jiang2025scimf}. Similarly, CytoBridge models stochastic cell-state transitions, unbalanced mass growth, and mean-field interactions via an unbalanced Schrödinger bridge \cite{zhang2025cytobridge}. However, those methods were mainly formulated for time-series scRNA-seq data. Therefore, their interaction kernels operate on the transcriptomic space and cannot leverage physical proximity to distinguish short-range microenvironmental signaling from long-range population effects.

Cell–cell interactions coordinate collective migration and cell-state transitions during development, homeostasis, and repair, allowing cell populations to self-organize and maintain complex tissue structures \cite{falco2025aggregation}. Mean field and nonlocal continuum theories provide an operator-theoretic foundation for resolving these multiscale interaction dynamics, representing collective interactions through spatial kernels integrated against the population density \cite{pal2025nonlocal,bernoff2016,falco2025aggregation}. Under suitable regularity, kernel-moment, and length-scale assumptions, long-wave expansions replace these nonlocal convolutions with local differential operators, such as the degenerate Cahn–Hilliard approximation for biological aggregation \cite{bernoff2016} and the local fourth-order aggregation–diffusion equation for differential cell adhesion \cite{falco2025aggregation}. This mathematical structure motivates separating short-range neighborhood effects from broader tissue context in expression dynamics: we represent long-range population cues through a mean-field attention integral outside radius R, and short-range microenvironmental cues through an anisotropic, geometry-aware graph-difference operator over pairwise expression contrasts within  R.

Here, we present \textbf{stMIND} (\textbf{S}patiotemporal \textbf{M}ultiscale \textbf{I}nteraction \textbf{N}eural \textbf{D}ynamics), a continuous-time generative model that extends our interacting mean-field neural SDE framework \cite{jiang2025scimf} to spatiotemporal transcriptomics by synthesizing McKean--Vlasov dynamics, graphon mean-field theory, and biological aggregation models \cite{mckean1966,fabian2023sparse,bayraktar2023graphon,bernoff2016}. Starting from unmatched spatial transcriptomic snapshots, stMIND jointly evolves single-cell gene expression and tissue coordinates. Crucially, stMIND decomposes the transcriptomic drift of each focal cell into a cell-intrinsic neural network and two interaction operators defined over dynamically updated, spatially disjoint context sets. The \emph{nonlocal operator} employs mass-weighted attention to aggregate signals from cells outside a physical radius $R$, acting as a learned integral operator over the exterior empirical measure. The \emph{local operator} acts over the co-evolving radius graph within distance $R$, aggregating relative expression differences modulated by pairwise spatial displacements—analogous to a state-dependent graph Laplacian or difference operator. To ensure structural identifiability among the additive drift components, stMIND incorporates an orthogonality regularizer alongside a dynamic optimal-transport objective in joint expression--position space.

We validate stMIND on time-series spatial transcriptomics atlases of axolotl telencephalon regeneration \cite{wei2022} and zebrafish embryogenesis \cite{liu2022zesta}. Across held-out forecasting benchmarks, stMIND achieves state-of-the-art expression accuracy at short horizons and remains highly competitive at long horizons. Systematic ablations in axolotl regeneration demonstrate that combining disjoint local and nonlocal operators yields the most accurate reconstruction of local cellular microenvironments. Finally, by interrogating the learned multiscale vector fields, stMIND pinpoints a critical regenerative transition window in the axolotl dorsal telencephalon and accurately resolves stage-specific developmental trajectories in zebrafish.
\par\endgroup

\section{Method}

\subsection{Overview of stMIND}

stMIND models a time-series of spatial transcriptomic sections as unmatched snapshots sampled from an evolving, proliferating, and migrating cell population (Fig.~1a). Each cell $i$ observed at an initial stage provides an initial gene-expression vector $x_i^t \in \mathbb{R}^d$, a physical tissue coordinate $s_i^t \in \mathbb{R}^{d_s}$, and a positive mass weight $m_i^t \in \mathbb{R}_+$ representing its contribution to the population distribution. To describe how its expression state evolves, stMIND uses an
interacting stochastic differential equation:
\begin{equation}
\mathrm{d}x_i^t
=
f_i^x(X^t,S^t,m^t)\,\mathrm{d}t
+
\sigma_x\,\mathrm{d}W_i^x,
\end{equation}
Here $X^t$, $S^t$, and $m^t$ collect the modeled expression states,
positions, and population weights at time $t$. The velocity vector, or the drift term of SDE $f_i^x$ determines the modeled direction of expression change and
can depend on both the focal cell and its cellular context.

To model cell migration and proliferation, we formulate tissue-position and
log-weight dynamics using separate learned fields, as in stVCR
\cite{peng2026}. Tissue positions evolve according to
\begin{equation}
\mathrm{d}s_i^t
=
f_i^s(x_i^t,s_i^t,t)\,\mathrm{d}t,
\label{eq:spatial-dynamics}
\end{equation}
where $f_i^s$ is the learned spatial velocity field. Population
change is represented by positive cell weights, initialized at
$m_i^{t_0}=1$, whose log-growth rates satisfy
\begin{equation}
\mathrm{d}\log m_i^t
=
r_i(x_i^t,s_i^t,t)\,\mathrm{d}t,
\label{eq:weight-dynamics}
\end{equation}
where $r_i$ is the learned log-weight growth-rate field.

To resolve cellular context across distinct spatial scales, stMIND decomposes the transcriptomic drift $f_i^x$ of focal cell $i$ into three additive vector fields:
\begin{equation}
f_i^x(X^t, S^t, m^t) = f_{\text{intrinsic}}(x_i^t) + f_{\text{nonlocal}, i}(X^t, m^t) + f_{\text{local}, i}(X^t, S^t, m^t).
\end{equation}
Here, $f_{\text{intrinsic}}$ depends solely on the focal cell's internal transcriptomic state $x_i^t$. The interaction terms $f_{\text{local}, i}$ and $f_{\text{nonlocal}, i}$ aggregate intercellular cues from two mutually exclusive sets of context cells partitioned by a physical interaction radius $R$, which are dynamically updated as cells migrate along a learned spatial velocity field (Fig.~1b). The choice of $R$ was informed by GITIII \cite{xiao2026gitiii}; the radius-calibration procedure and fitting-range-specific parameter values are provided in Section 2 of the Supplementary Material.

The \textbf{nonlocal operator} $f_{\text{nonlocal}, i}$ captures long-range population influences from context cells residing outside the focal cell's immediate neighborhood ($\|s_j^t - s_i^t\|_2 \ge R$). Extending the mean-field attention mechanism of scIMF \cite{jiang2025scimf}, the focal cell's expression state generates a query, while eligible distant context cells supply expression-derived keys and values weighted by their evolving masses $m_j^t$. At the continuum level, this mass-normalized attention acts as a learned integral operator over the exterior empirical measure.

Conversely, the \textbf{local operator} $f_{\text{local}, i}$ models short-range microenvironmental interactions over a dynamic radius graph ($\|s_j^t - s_i^t\|_2 < R$). Rather than aggregating raw expression states, the local operator aggregates pairwise expression differences $(x_j^t - x_i^t)$ between neighboring context cells and the focal cell, with attention keys modulated by a learned spatial encoding of their relative displacement $(s_j^t - s_i^t)$. Mathematically, this formulation endows $f_{\text{local}, i}$ with the structure of an anisotropic, state-dependent graph difference operator whose response vanishes when a neighborhood is transcriptionally homogeneous and adapts dynamically as local tissue geometry reorganizes.

stMIND is trained end-to-end by minimizing a relaxed dynamic least-action objective that couples trajectory kinetic energy with optimal transport and spatial occupancy matching at observation time points. Because supervising only marginal snapshots leaves an additive decomposition under-constrained, we introduce an endpoint orthogonality penalty that discourages directional collinearity between $f_{\text{intrinsic}}$ and each interaction term, ensuring a parsimonious and identifiable separation of intrinsic and extrinsic dynamics (Fig.~1d,e).

\begin{figure}[!htbp]

\centering

\includegraphics[width=\linewidth,height=0.55\textheight,keepaspectratio]{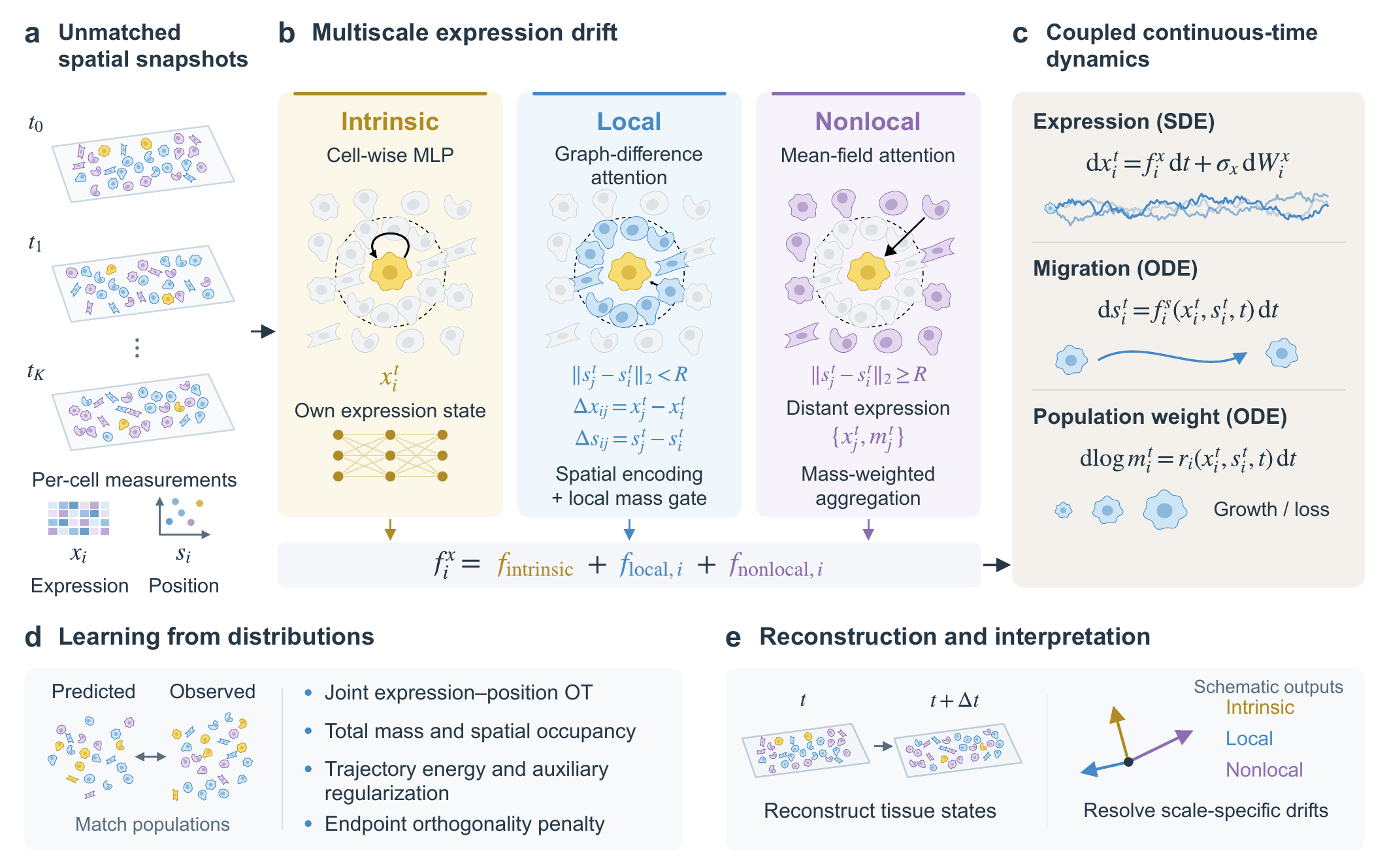}

\caption{\textbf{Overview of stMIND.} (a) Unmatched spatial-transcriptomic snapshots provide paired gene-expression states and tissue coordinates for individual cells
at discrete observation times. (b) stMIND decomposes expression drift into three additive components (c) The summed expression drift drives a stochastic differential
equation, while separate learned ordinary differential equations
govern tissue positions and log particle weights. (d) Training matches predicted and observed populations using a
balanced Sinkhorn divergence in joint expression--position space,
separate total-mass and spatial-occupancy supervision, and trajectory
regularization. An endpoint orthogonality penalty discourages overlap
between intrinsic and interaction-driven drifts through
mass-weighted absolute inner products.
(e) The fitted model supports reconstruction and forecasting of
tissue states and analysis of intrinsic, local, and nonlocal
expression-drift contributions.}\label{fig:overview}

\end{figure}

\subsection{Neural network architecture}

\label{sec:model}

For each focal cell $i$ at time $t$, we represent its expression
drift as the sum of three learned operators: a cell-intrinsic
operator $f_{\mathrm{intrinsic}}$, a nonlocal interaction operator
$f_{\mathrm{nonlocal},i}$, and a local graph-difference operator
$f_{\mathrm{local},i}$. The two interaction operators act on
disjoint sets of context cells defined by the spatial radius $R$:
a local neighborhood
$\mathcal N_R(i,t)=
\{j\ne i:\|s_j^t-s_i^t\|_2<R\}$
and a nonlocal exterior
$\mathcal N_R^c(i,t)=
\{j\ne i:\|s_j^t-s_i^t\|_2\ge R\}$.
These context sets are updated as cell positions $S^t$ evolve.

\paragraph{Intrinsic operator.}
The intrinsic operator
$f_{\mathrm{intrinsic}}:\mathbb R^{d_x}\to\mathbb R^{d_x}$
is parameterized by a cell-wise multilayer perceptron (MLP)
shared across all modeled cells. It takes the focal cell's
expression state $x_i^t$ as input and produces its intrinsic
expression-drift contribution in the same expression space.

\paragraph{Nonlocal interaction operator.}
For a single attention head of the nonlocal operator, the focal cell forms query $\tilde{q}_i = \tilde{W}_Q x_i^t$, while each distant cell $j \in \mathcal{N}_R^c(i, t)$ provides key $\tilde{k}_j = \tilde{W}_K x_j^t$ and value $\tilde{v}_j = \tilde{W}_V x_j^t$. Applying a spatial mask to exclude $\{i\} \cup \mathcal{N}_R(i, t)$ and incorporating context-cell masses $m_j^t$, the nonlocal drift contribution aggregates distant signals as:
\begin{equation}
f_{\text{nonlocal}, i}(X^t, m^t) = \tilde{W}_O \sum_{j \in \mathcal{N}_R^c(i, t)} \alpha_{ij}^{\text{nonlocal}}\, \tilde{v}_j, \quad \text{where} \quad \alpha_{ij}^{\text{nonlocal}} = \frac{m_j^t \exp(\langle \tilde{q}_i, \tilde{k}_j \rangle / \sqrt{d_k})}{\sum_{\ell \in \mathcal{N}_R^c(i, t)} m_\ell^t \exp(\langle \tilde{q}_i, \tilde{k}_\ell \rangle / \sqrt{d_k})}.
\end{equation}

\paragraph{Local graph-difference operator.}
In contrast, the local operator acts strictly on $\mathcal{N}_R(i, t)$ and couples relative spatial geometry with local transcriptomic gradients. For each head, the query, spatially modulated key, and difference-based value are defined as:
\begin{equation}
q_i = W_Q x_i^t, \qquad k_{ij} = W_K x_j^t + E_\theta(s_j^t - s_i^t), \qquad v_{ij} = W_V(x_j^t - x_i^t), \quad j \in \mathcal{N}_R(i, t),
\label{eq:local_qkv}
\end{equation}
where $E_\theta : \mathbb{R}^{d_s} \to \mathbb{R}^{d_k}$ is a learned relative-position encoding network. The attention-weighted values are projected to the local drift via a neighborhood-mass gate $g(m_{\mathcal{N}_R(i, t)}^t)$ controlled by the local mass $m_{\mathcal{N}_R(i, t)}^t = \sum_{j \in \mathcal{N}_R(i, t)} m_j^t$:
\begin{equation}
f_{\text{local}, i}(X^t, S^t, m^t) = g\big(m_{\mathcal{N}_R(i, t)}^t\big) W_O \sum_{j \in \mathcal{N}_R(i, t)} \alpha_{ij}^{\text{local}}\, v_{ij}, \quad \text{where} \quad \alpha_{ij}^{\text{local}} = \frac{m_j^t \exp(\langle q_i, k_{ij} \rangle / \sqrt{d_k})}{\sum_{\ell \in \mathcal{N}_R(i, t)} m_\ell^t \exp(\langle q_i, k_{i\ell} \rangle / \sqrt{d_k})}.
\end{equation}
The nonlocal and local operators maintain completely independent parameter sets.
The architecture of the intrinsic, nonlocal, and local expression-drift branches is illustrated in Supplementary Fig.~S1, with the drift decomposition and interaction operators defined in Supplementary Eqs.~(S3)–(S11).

In addition to these three expression-drift terms, separate
multilayer perceptrons (MLPs) predict spatial velocity and
log-weight growth rate from each cell's expression, position, and
time. During numerical integration, local radius graphs and complementary nonlocal masks are updated using the current cell positions, while a sampled context-cell pool reduces the cost of nonlocal attention. Detailed network configurations and implementation settings are provided in Section 2 of the Supplementary Material.

\subsection{Training from unmatched snapshots}

To learn the coupled dynamics from $K$ unmatched target snapshots $\{t_1, \dots, t_K\}$, stMIND minimizes a relaxed least-action and distribution-matching objective:
\begin{equation}
\mathcal{J} = \mathbb{E}\left[ \lambda_E \mathcal{L}_{\text{energy}} + \frac{1}{K}\sum_{k=1}^K \left( \lambda_{\text{OT}}\mathcal{L}_{\text{OT}, k} + \lambda_M \mathcal{L}_{M, k} + \lambda_H \mathcal{L}_{H, k} \right) + \mathcal{L}_{\text{aux}} + \lambda_\perp(e)\mathcal{L}_\perp \right].
\label{eq:loss}
\end{equation}
Here, $\mathcal{L}_{\text{energy}}$ penalizes cumulative expression drift norm and spatial kinetic energy along simulated trajectories (following Benamou--Brenier dynamic optimal transport); $\mathcal{L}_{\text{OT}, k}$ is an entropic optimal-transport loss matching predicted and observed joint expression--position distributions at snapshot $t_k$; and $\mathcal{L}_{M, k}$ and $\mathcal{L}_{H, k}$ supervise total population mass and spatial tissue occupancy, respectively. $\mathcal{L}_{\text{aux}}$ collects regularizers on growth rates, spatial support, and local drift magnitude. To address the identifiability issue in the additive drift
decomposition, we introduce an orthogonality penalty
$\mathcal L_{\perp}$, which discourages overlap between intrinsic
and interaction-driven drifts. Specifically, it penalizes the
absolute inner products between $f_{\mathrm{intrinsic}}$ and each
of $f_{\mathrm{nonlocal}}$ and $f_{\mathrm{local}}$, using normalized
population weights and averaging over the noninitial observation
endpoints. This penalty depends on both the directions and
magnitudes of the drift components. Complete mathematical definitions of the objective and all loss terms are provided in Section 1.3 of the Supplementary Material. Hyperparameters, optimization settings, and numerical integration procedures are detailed in Section 2 of the Supplementary Material.

\section{Results}

\begingroup\color{black}
\subsection{stMIND accurately predicts gene expression and spatial distributions}

We first evaluated stMIND on two time-resolved spatial transcriptomics datasets, axolotl brain regeneration and zebrafish embryogenesis, against four baseline methods: two spatial-temporal frameworks, stVCR \cite{peng2026} and STORIES \cite{huizing2026stories}, and two non-spatial interacting mean-field methods, scIMF \cite{jiang2025scimf} and CytoBridge \cite{zhang2025cytobridge}. The axolotl dataset from the ARTISTA atlas records molecular and spatial changes in the telencephalon during regeneration after injury \cite{wei2022}. The zebrafish dataset from the ZESTA atlas captures changes in gene expression and tissue organization during early embryonic development \cite{liu2022zesta}. Data sources and preprocessing procedures are detailed in Section 3 of the Supplementary Material.

We defined short- and long-horizon extrapolation tasks for each dataset. Axolotl models fitted at 2, 5, 10, 15 and 20 days post-injury (DPI) were evaluated at 30 and 60 DPI. Zebrafish models fitted at 3.3, 5.25, 10 and 12 hours post-fertilization (hpf) were evaluated at 18 and 24 hpf.  We quantified marginal gene-expression accuracy using $W_1$ and $W_2$ Wasserstein distances in a shared reference embedding, joint transcriptomic--anatomical fidelity using Fused Gromov--Wasserstein (FGW) discrepancy, and spatial coherence of the predicted expression field using Moran's $I$. All metrics are reported as mean $\pm$ s.d.\ across five random seeds.

On the short-horizon prediction tasks, stMIND achieved the lowest mean gene-expression errors ($W_1$ and $W_2$) across all five methods in both biological systems (Fig.~2a,b). In axolotl regeneration at 30 DPI, stMIND attained $W_1 = 2.287$ and $W_2 = 2.386$ (Fig.~2a). Similarly, in zebrafish development at 18 hpf, stMIND achieved the best expression errors with $W_1 = 2.179$ and $W_2 = 2.265$ (Fig.~2b). At the long-horizon endpoints, stMIND ranked second overall in marginal expression accuracy: at 60 DPI in axolotl, stMIND achieved $W_1 = 3.042$ and $W_2 = 3.147$, trailing only its non-spatial predecessor scIMF ($W_1 = 2.501, W_2 = 2.588$); at 24 hpf in zebrafish, stMIND achieved $W_1 = 2.457$ and $W_2 = 2.618$, closely behind scIMF ($W_1 = 2.362, W_2 = 2.468$). This pattern reflects a natural trade-off at extended extrapolation horizons: whereas scIMF optimizes trajectories unconstrained in expression space alone, stMIND simultaneously enforces physical tissue transport and spatial organization constraints over long unobserved intervals.

Crucially, unlike scIMF, CytoBridge, and STORIES—which cannot natively evolve physical coordinates forward in time—stMIND jointly predicts evolving tissue geometry. Compared against stVCR (the only baseline capable of generating spatial coordinates), stMIND achieved substantially lower FGW discrepancy across every completed benchmark, demonstrating superior joint alignment of gene expression and anatomical structure (Fig.~2a,b). Furthermore, stMIND generated stronger positive spatial autocorrelation (Moran's $I$) than stVCR at both 30 and 60 DPI in axolotl, while stVCR exhibited higher Moran's $I$ at 18 hpf in zebrafish. At the 24-hpf zebrafish endpoint, stVCR failed to complete within the computational time limit, leaving stMIND as the only method capable of delivering full joint spatiotemporal predictions at scale.
\par\endgroup

\begin{figure}[!tbp]
 \centering
 \color{black}
 \includegraphics[width=\linewidth,height=0.55\textheight,keepaspectratio]{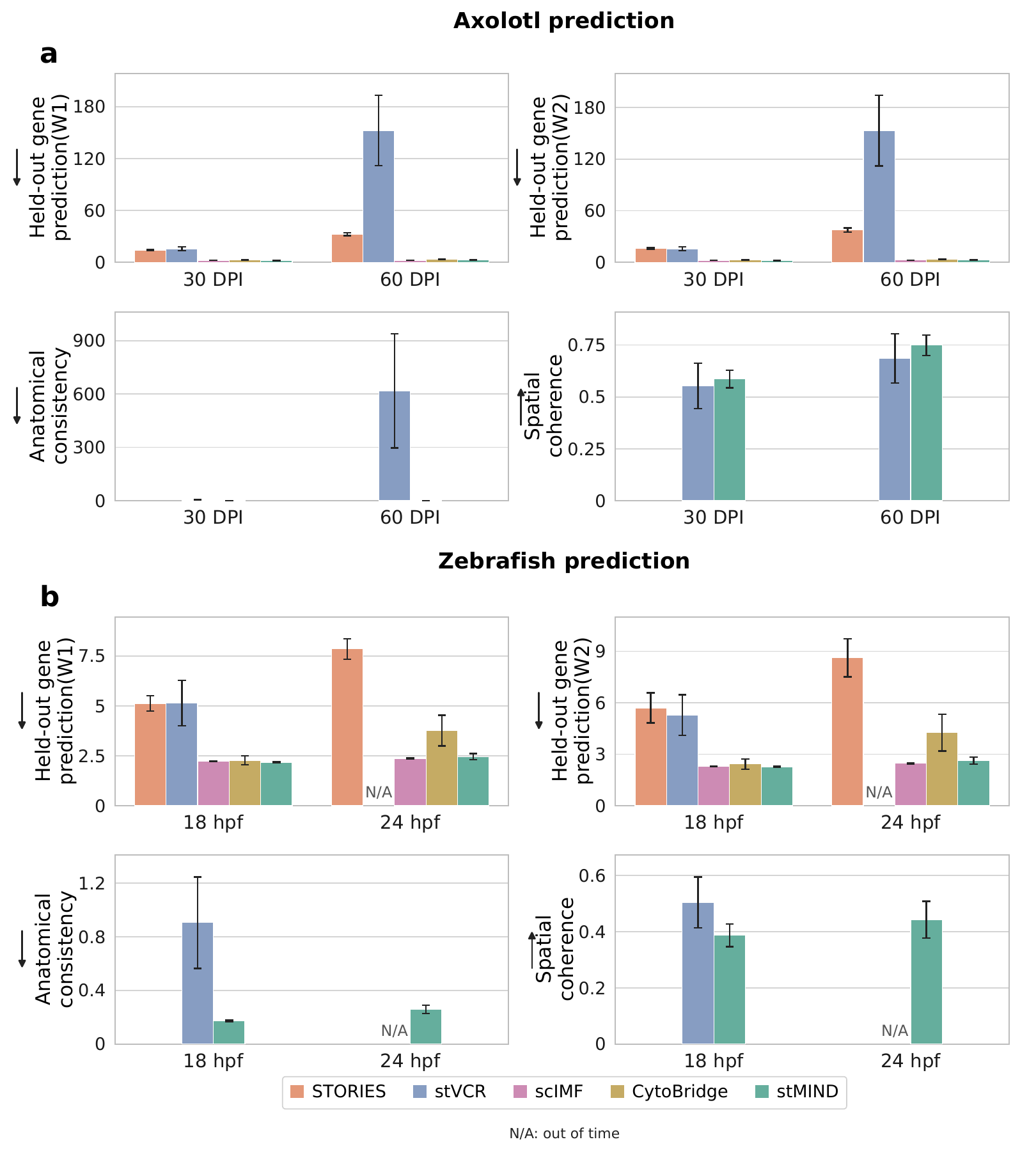}
 \captionsetup{font=normalsize,labelfont={bf,color=black},textfont={color=black}}
 \caption{\textbf{Performance evaluation of stMIND versus baseline methods on held-out prediction tasks.} Comparison of stMIND with stVCR, STORIES, scIMF and CytoBridge in axolotl brain regeneration and zebrafish embryogenesis. $W_1$ and $W_2$ measure held-out expression-distribution error. The anatomical-consistency panel shows FGW discrepancy. Spatial coherence is the predicted expression Moran's $I$. \textbf{(a)} Axolotl predictions at 30 and 60 days post-injury (DPI). \textbf{(b)} Zebrafish predictions at 18 and 24 hours post-fertilization (hpf). N/A denotes incomplete stVCR predictions at 24 hpf.}
 \label{fig:forecast_benchmark}
\end{figure}

\begin{figure}[!tbp]
 \centering
 \color{black}
 \includegraphics[width=\linewidth,height=0.55\textheight,keepaspectratio]{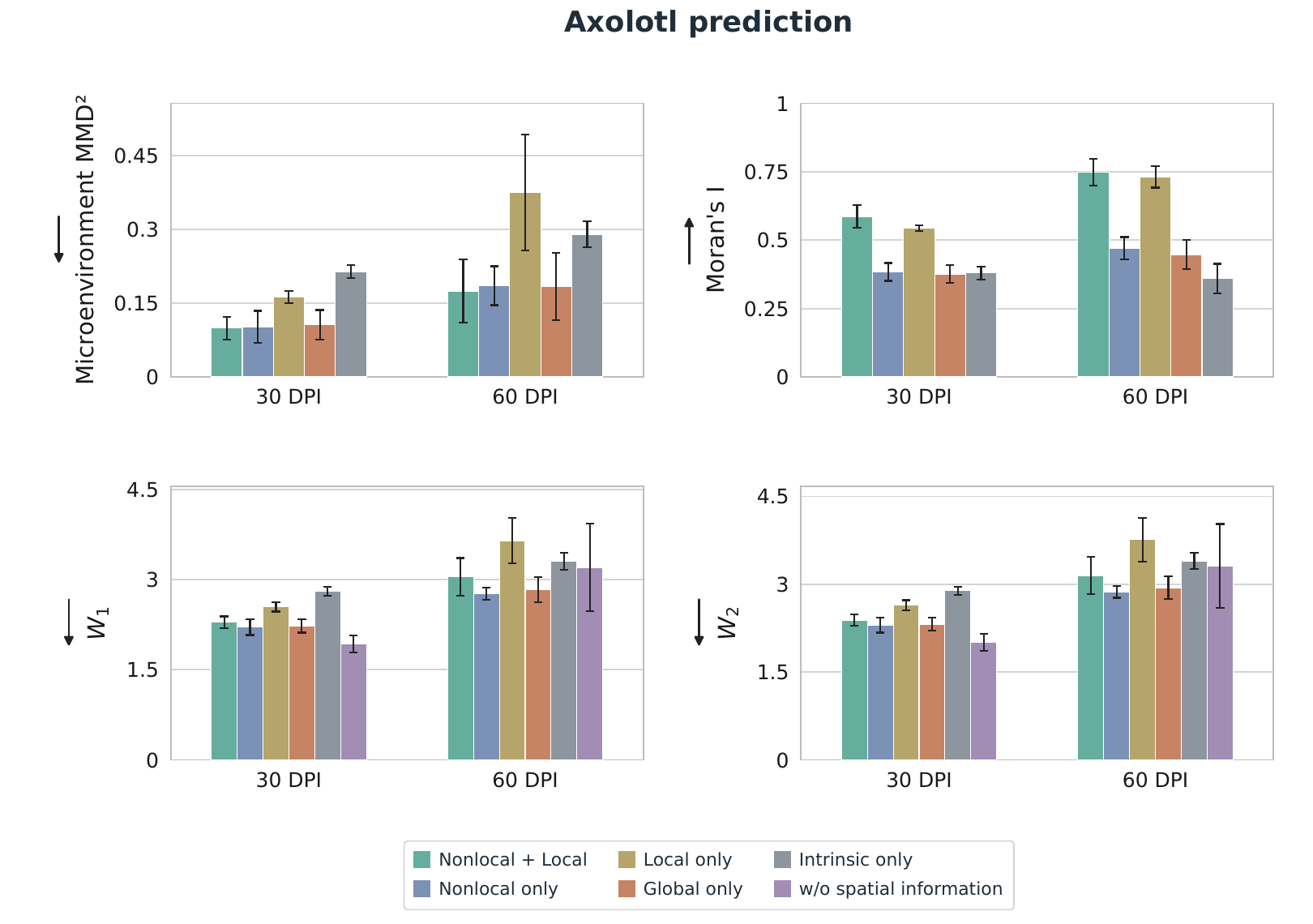}
 \captionsetup{font=normalsize,labelfont={bf,color=black},textfont={color=black}}
 \caption{\textbf{Ablation of stMIND's interaction terms and spatial information.} stMIND and its independently retrained variants were fitted at 2--20 days post-injury (DPI), with predictions evaluated at 30 and 60 DPI. Microenvironment MMD$^2$, predicted Moran's $I$, and gene-expression $W_1$ and $W_2$ are shown for Nonlocal + local (stMIND), Nonlocal only, Local only, Global only, Intrinsic only, and a model without spatial information.}
 \label{fig:interaction_ablation}
\end{figure}

\begingroup\color{black}
\subsection{stMIND's nonlocal and local terms improve spatial microenvironment prediction}

To dissect the functional contributions of local interactions, nonlocal interactions, and spatial geometry, we conducted a comprehensive six-way ablation study on the axolotl forecasting benchmark (trained on 2--20 DPI; evaluated at 30 and 60 DPI; Fig.~\ref{fig:interaction_ablation}). Alongside the full model (\emph{Nonlocal + local}), we evaluated five independently retrained variants across the same five seeds: (i) \emph{Nonlocal only}, which removes $f_{\text{local}}$ while preserving the spatial exclusion mask $\mathcal{N}_R^c(i,t)$; (ii) \emph{Local only}, which removes $f_{\text{nonlocal}}$; (iii) \emph{Global only}, which removes both $f_{\text{local}}$ and the radius mask, collapsing the interaction term into unmasked global population attention; (iv) \emph{Intrinsic only}, which removes all intercellular interactions while retaining spatial velocity $f^s_i$ and growth $r_i$; and (v) \emph{w/o spatial information}, which removes tissue coordinates, spatial dynamics, and spatial supervision entirely.

Among the five spatially informed variants, the full stMIND model (\emph{Nonlocal + local}) achieved the lowest mean microenvironment $\text{MMD}^2$ at both 30 and 60 DPI (Fig.~\ref{fig:interaction_ablation}), a joint distributional metric that evaluates how accurately a model reconstructs focal cell expression alongside the transcriptomic and cell-type composition of its surrounding physical niche (see Supplementary Eq. (S31) for its detailed definition). While \emph{Nonlocal only} and \emph{Global only} improved upon \emph{Intrinsic only}, neither matched the microenvironment fidelity of the full multiscale decomposition.

Evaluating spatial coherence within the interaction radius $R$ further highlighted the necessity of the local graph-difference operator. Full stMIND achieved the highest predicted Moran's $I$ at both 30 DPI ($0.586$) and 60 DPI ($0.749$), substantially outperforming \emph{Global only} ($0.376$ and $0.447$, respectively; Fig.~\ref{fig:interaction_ablation}). Notably, \emph{Local only} also maintained strong spatial autocorrelation, confirming that local difference aggregation actively preserves spatially coherent expression domains as cells migrate.

In marginal gene-expression metrics ($W_1$ and $W_2$), the unconstrained \emph{w/o spatial information} variant yielded the lowest error at 30 DPI, while \emph{Nonlocal only} achieved the lowest error at 60 DPI (Fig.~\ref{fig:interaction_ablation}). Among spatial variants, full stMIND consistently outperformed \emph{Local only} and \emph{Intrinsic only}, while trading a modest margin in global $W_1/W_2$ relative to \emph{Nonlocal only} in exchange for superior microenvironment reconstruction ($\text{MMD}^2$) and spatial coherence (Moran's $I$). Thus, combining spatially disjoint local and nonlocal operators is essential for faithfully reconstructing structured multicellular microenvironments.
\par\endgroup

\begin{figure}[!htbp]
 \centering
 \color{black}
 \includegraphics[width=\linewidth,height=0.74\textheight,keepaspectratio]{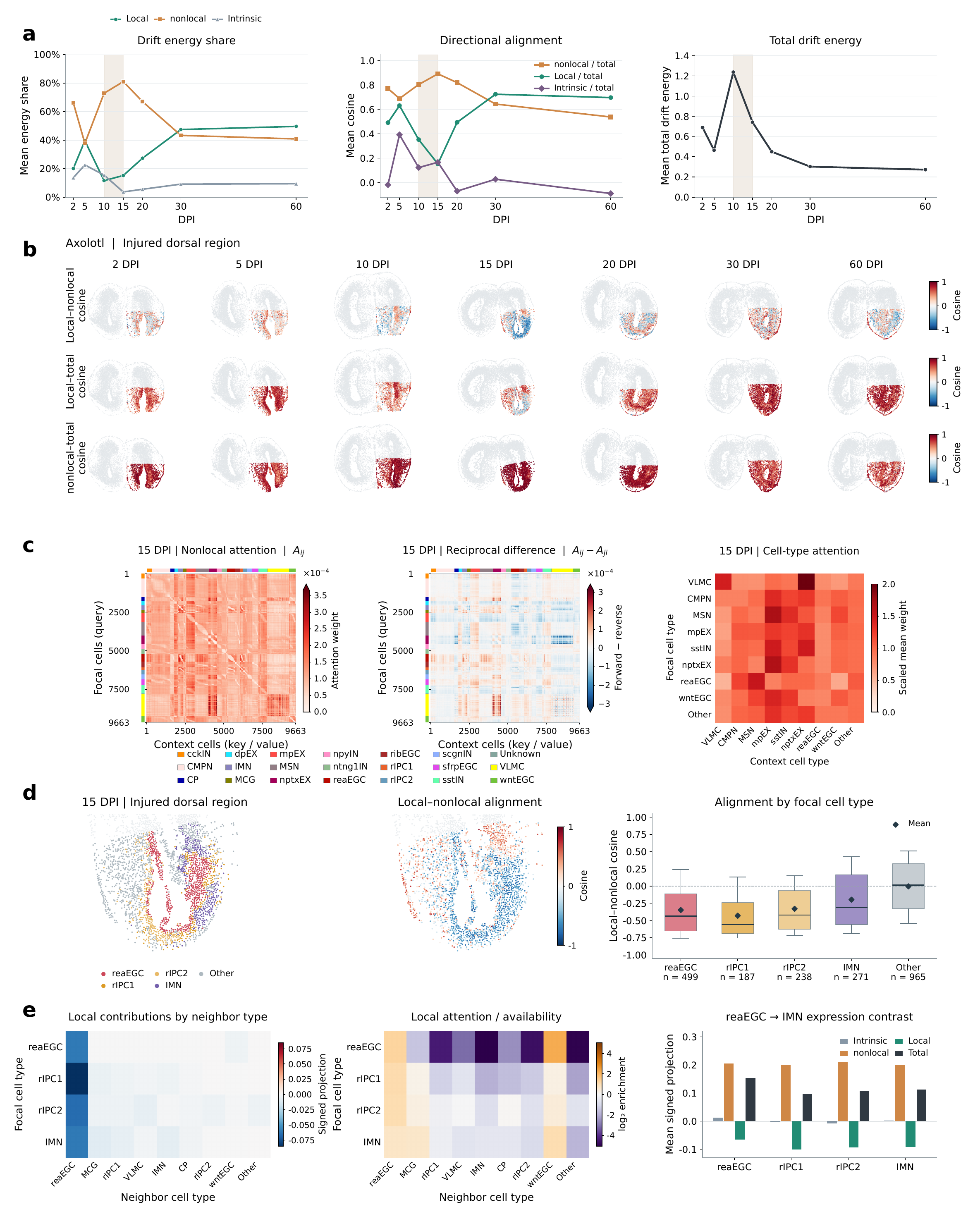}
 \captionsetup{font=normalsize,labelfont={bf,color=black},textfont={color=black}}
 \caption{\textbf{Multiscale expression-drift energy and cell-context contributions during axolotl regeneration.} \textbf{(a)} Stage-wise energy shares of the intrinsic, nonlocal and local drift terms, their directional alignment with total drift, and total drift energy in the injured dorsal region. \textbf{(b)} Spatial maps of local--nonlocal, local--total and nonlocal--total alignment. \textbf{(c)} Nonlocal attention, reciprocal differences and cell-type-pair means at 15 DPI. \textbf{(d)} Cell types and local--nonlocal alignment within the injured dorsal region at 15 DPI. \textbf{(e)} Neighbor-type local contributions, attention enrichment and drift projections along the reaEGC--IMN expression contrast.}
 \label{fig:axolotl_transition}
\end{figure}

\begingroup\color{black}
\subsection{stMIND identifies a critical regenerative transition through multiscale drift-energy reorganization}

To investigate how multiscale cellular interactions reorganize during tissue repair, we trained stMIND across the complete axolotl regeneration time course (2 to 60 DPI) and analyzed the learned drift vector fields within the injured dorsal telencephalon, which we defined as the region of interest (ROI). We defined the \emph{expression-drift energy} of each component as its squared Euclidean norm ($\|f\|^2$), providing a quantitative measure of kinetic energy across stages. This analysis uncovered a critical regenerative transition window between 10 and 15 DPI, marked by an acute surge in total expression-drift energy followed by a sharp reorganization between nonlocal and local interaction scales (Fig.~\ref{fig:axolotl_transition}a,b). Furthermore, the expression-drift energy stabilizes into an equilibrium state by 30 and 60 DPI (Fig.~\ref{fig:axolotl_transition}a).

Specifically, expression-drift energy peaked at 10 DPI and subsequently declined, revealing pronounced expression-dynamical activity during the intermediate phase of regeneration (Fig.~\ref{fig:axolotl_transition}a, Supplementary Figs. S2 and S3). Following this peak, the nonlocal share of summed component energies reached its maximum at 15 DPI, accounting for approximately 81\%. At the same stage, nonlocal drift exhibited its strongest alignment with total drift, whereas local--total alignment was weakest. Spatial analysis further revealed the strongest local--nonlocal opposition at 15 DPI, with negative alignment in approximately 65\% of cells within the ROI (Fig.~\ref{fig:axolotl_transition}b). This transition window therefore encompassed two successive dynamical features: elevated total drift energy at 10 DPI, followed by nonlocal dominance and local directional counterbalance at 15 DPI. At 30 and 60 DPI, total drift energy had decreased and the relative energy shares of the local and nonlocal components had become more balanced, suggesting that the multiscale dynamics had largely stabilized.

To investigate how nonlocal information was organized at this transition, we examined nonlocal attention at 15 DPI (Fig.~\ref{fig:axolotl_transition}c). The cell-level attention matrix and its reciprocal difference revealed structured asymmetry in the weights assigned between cell pairs, while cell-type summaries showed heterogeneous patterns of context weighting. Stage-dependent patterns of nonlocal attention and nonreciprocity were also observed across the full regeneration series
(Supplementary Figs.~S4 and~S5). These results characterize nonlocal aggregation as a focal-cell-dependent process in which the two directions of a cell pair can receive different weights.

Within the ROI, local--nonlocal directional balance further exhibited cell-type dependence (Fig.~\ref{fig:axolotl_transition}d, Supplementary Fig.~S6). Opposing directions were particularly pronounced in reactive ependymoglial cells (reaEGCs) and reactive intermediate progenitor cell (rIPC) subtypes, with the strongest opposition in rIPC1. Immature neurons (IMNs) also showed negative mean alignment, whereas the pooled remaining annotations had a mean close to zero. These findings localized the mid-regeneration dynamical reorganization to specific cell populations, linking distinct cellular states to different combinations of multiscale contributions.

Decomposition of local contributions further connected this pattern to reaEGC-associated neighborhoods (Fig.~\ref{fig:axolotl_transition}e). Across all four focal-cell groups, reaEGC neighbors received a larger attention fraction than their availability among neighboring cells. Along a fixed reaEGC-to-IMN expression-contrast axis, reaEGC neighbors supplied the largest absolute group-mean local contribution. Nonlocal projections were positive, whereas local projections were negative, partially offsetting the nonlocal contribution while leaving a positive total projection. Thus, local neighborhoods contributed to the resulting expression drift through directional counterbalance. Together, stMIND identified a key regenerative window in which elevated total drift energy was followed by nonlocal dominance and cell-type-dependent local counterbalance, and linked this stage-specific dynamical pattern to reaEGC-associated neighborhoods. These findings provide a testable hypothesis for how cellular context at different spatial scales contributes to cell-state reorganization during regeneration.
\par\endgroup

\begin{figure}[!htbp]
 \centering
 \color{black}
 \includegraphics[width=\linewidth,height=0.55\textheight,keepaspectratio]{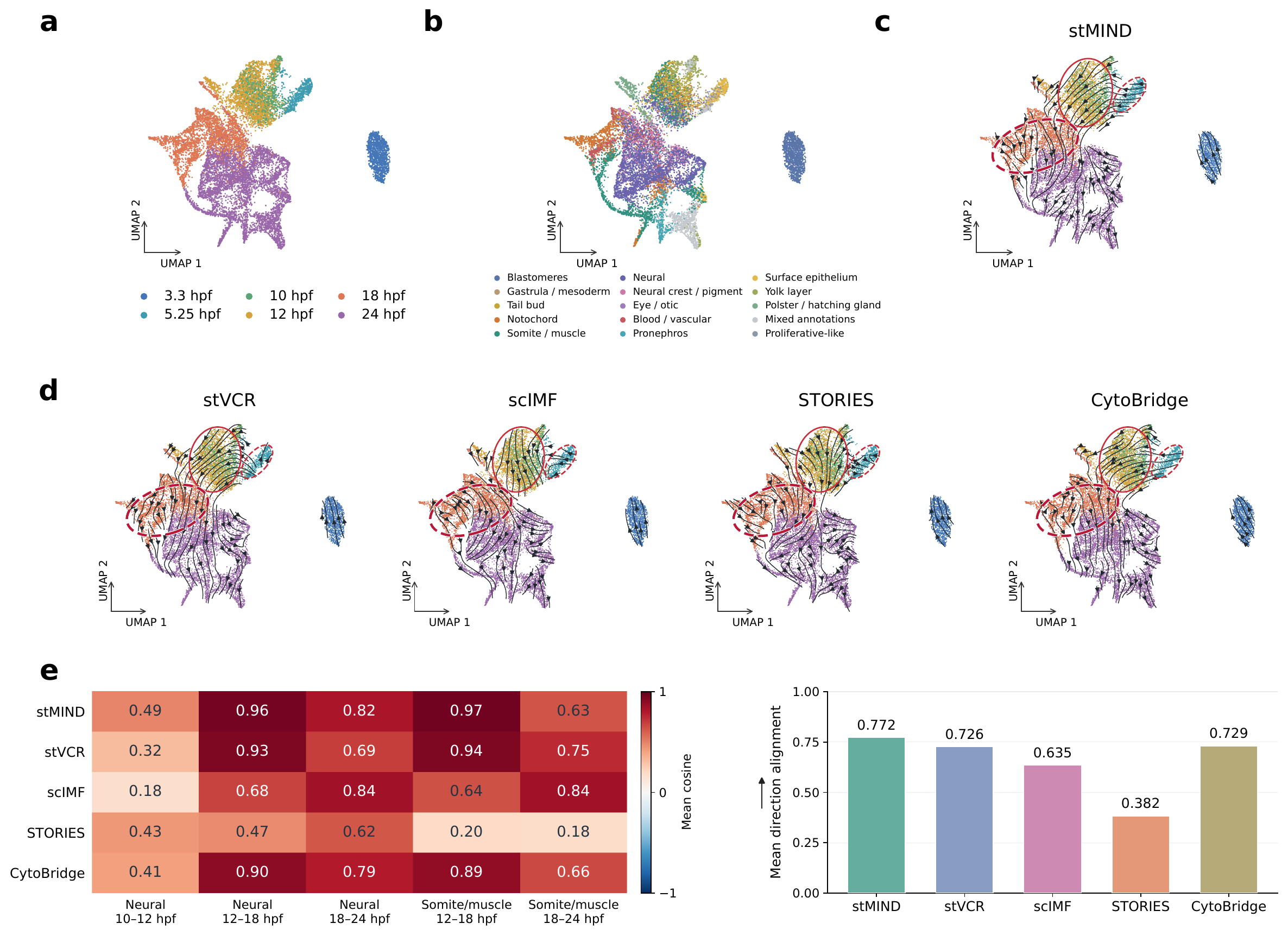}
 \captionsetup{font=normalsize,labelfont={bf,color=black},textfont={color=black}}
 \caption{\textbf{Stage-dependent expression dynamics during zebrafish development.} Comparison of inferred expression drift on the same observed cells. \textbf{(a)} UMAP colored by developmental stage. \textbf{(b)} The same UMAP colored by broad cell group. \textbf{(c)} stMIND expression-drift directions on the shared UMAP. \textbf{(d)} Corresponding directions for stVCR, scIMF, STORIES, and CytoBridge. \textbf{(e)} Mean cosine alignment with a within-group adjacent-stage reference for neural and somite/muscle populations (left), and the equal-weight mean across the five displayed comparisons (right).}
 \label{fig:zebrafish_velocity}
\end{figure}

\begingroup\color{black}
\subsection{stMIND resolves stage-dependent expression dynamics during zebrafish development}

We next examined whether stMIND captures the changing expression dynamics of zebrafish development. In the shared expression atlas, early blastomeres separated from later populations, while neural, axial, and muscle-related states occupied partly overlapping regions (Fig.~\ref{fig:zebrafish_velocity}a,b). We first examined stMIND's projected expression drift on this atlas (Fig.~\ref{fig:zebrafish_velocity}c), then compared it with the corresponding vector fields inferred by stVCR, scIMF, STORIES, and CytoBridge on the same observed cells (Fig.~\ref{fig:zebrafish_velocity}d, Supplementary Fig.~S7). The highlighted regions are enriched for 5.25, 10--12, and 18 hpf cells, allowing the projected directions to be read against successive developmental stages. stMIND broadly follows the progression from the 5.25-hpf region toward the 10--12-hpf states and onward toward later populations. CytoBridge shows a similar early direction, whereas stVCR and STORIES less clearly follow the shift from 5.25 to 10 hpf. stVCR more closely follows the middle-stage shift toward 18 hpf, while scIMF and STORIES show weaker correspondence there. For the later shift toward 24 hpf, scIMF follows the observed progression particularly well, while stVCR and CytoBridge vary between neural and somite/muscle populations.

To assess these visual patterns, we compared inferred drift with within-group changes toward the next observed stage in neural and somite/muscle populations (Fig.~\ref{fig:zebrafish_velocity}e, left). stMIND showed positive directional agreement in every displayed comparison and the highest average agreement (Fig.~\ref{fig:zebrafish_velocity}e, right). stVCR and CytoBridge were close at some middle-stage transitions, whereas scIMF performed well at the latest transitions. STORIES showed the weakest overall agreement, especially in somite/muscle cells.
\par\endgroup

\section{Discussion}

\begingroup\color{black}
Dissecting how tissues develop and regenerate requires disentangling cell-intrinsic regulatory programs from short-range microenvironmental signaling and long-range population cues. These interactions co-evolve as cells proliferate and migrate, yet spatial transcriptomics provides unmatched snapshots through destructive sampling. Reconstructing the underlying multiscale dynamics therefore remains challenging. Extending our interacting mean-field neural SDE framework (scIMF \cite{jiang2025scimf}) into physical space, stMIND bridges this gap by jointly evolving gene expression, spatial coordinates, and cellular mass while decomposing transcriptomic drift into intrinsic, local graph-difference, and nonlocal mean-field operators.

Separating local and nonlocal context yielded complementary insights. In axolotl ablations, the full model more accurately reconstructed local microenvironment distributions than other spatially informed variants while retaining competitive gene-expression prediction. In the full-series axolotl analysis, a peak in total expression-drift energy preceded nonlocal dominance and cell-type-dependent local counterbalance; the local response also preferentially drew on reaEGC neighbors. Nonreciprocal attention further shows that contextual dependence
is directional in the learned model: two cells can assign
different importance to one another. Combined with cell-type-dependent
local counterbalance, this provides a way to examine how focal-cell
state and spatial context jointly shape predicted expression change. In zebrafish, inferred expression drifts captured stage-dependent developmental progression. Together, these findings show how distinguishing interaction ranges supports both tissue prediction and analysis of coordinated molecular and anatomical change.

The multiscale decomposition also makes the relationship between
drift magnitude and direction explicit. A component with a smaller
energy share can still influence total expression change through
its orientation relative to the other components. Energy shares
quantify relative squared magnitudes, whereas directional alignment
and signed projections describe reinforcement and counterbalance.
The axolotl analysis illustrates why these readouts are
complementary: nonlocal dominance coexists with cell-type-dependent
local responses. This combination allows localized, state-dependent
responses to be examined even when nonlocal contributions dominate
the component energies.

The prominence of reaEGC-associated neighborhoods connects these
dynamical patterns to the established regenerative role of
ependymoglial cells. Targeted EGC depletion has been shown to block
axolotl brain regeneration \cite{fu2026egc}. In stMIND, reaEGC
neighbors received attention exceeding their neighborhood
availability and contributed to local counterbalance along the
reaEGC-to-IMN expression contrast. These observations motivate
the hypothesis that reactive glial neighborhoods help coordinate
regenerative cell-state transitions. Spatial profiling combined
with targeted perturbations could test whether changing
reaEGC-associated context alters the expression responses of
nearby progenitor and immature neuronal populations.

Several computational and modeling directions merit future
investigation. First, full pairwise nonlocal attention incurs $\mathcal{O}(N^2)$
computational cost per integration step, while repeatedly updating
local radius graphs adds further overhead during continuous SDE
rollouts. We currently reduce the attention cost through mini-batch
sampling of context cells, but this subsampling may omit informative
long-range cellular context. TissueFormer addresses quadratic
attention costs through a whole-slide attention mechanism with
linear complexity in the number of cells \cite{wu2026tissueformer}.
Adapting a related strategy to stMIND's evolving spatial masks and
mass-weighted aggregation could preserve broader nonlocal context
at lower computational cost and reduce reliance on context
subsampling. Second, integrating
curated ligand--receptor priors \cite{turei2021} into the query--key
projections of local and nonlocal attention, together with
gene-resolved features, could help relate learned expression-drift
contributions to candidate molecular signaling pathways. We plan
to investigate scalable attention and biologically informed
interaction terms to extend stMIND to larger tissue atlases and
generate more directly testable signaling hypotheses.
\par\endgroup
\section{Acknowledgements}

This work was supported by the Strategic Priority Research
Program of the Chinese Academy of Sciences
(grant no.~XDB1350203) and the National Key Research and
Development Program of China (grant no.~2022YFA1004801).
\clearpage

\bibliographystyle{unsrtnat}
\bibliography{references}

\end{document}